# Charge transfer induced cubic gauche nitrogen from azides

Tingting Ye[1#], Yuxuan Xu[1,2#], Guo Chen[1,2], Ming Li[1,2], Liangfei Wu[1], Jie Zhang[1,2], Junfeng Ding[1,2*] & Xianlong Wang[1,2*]

[1]Key Laboratory of Materials Physics, Institute of Solid State Physics, HFIPS, Chinese Academy of Sciences, Hefei 230031, China;

[2]University of Science and Technology of China, Hefei 230026, China;

*Corresponding author. Email: junfengding@issp.ac.cn; xlwang@theory.issp.ac.cn

#These authors contributed equally to this work.

# ABSTRACT

Cubic gauche nitrogen (cg-N) with a three-dimensional network of N-N single bonds attracted lots of attentions in last decades, since it theoretically has five times larger energy than TNT. While, extreme environments of high pressure or plasma treatment were required in traditional routes. Quite recently, in vacuum or protective gas environments, a one-step synthesis route relying solely on heating is reported giving the highest cg-N content. However, corresponding mechanism is missing, which hinders the improvement of yield and the development of simpler methods. Here, by treating $KN_3$ in different gas environments, we find that moisture can prevent the transition from $KN_3$ to cg-N. In a dry air environment at 260 ~ 300°C, $KN_3$ decomposes into K and $N_2$, and charge transfer from K to $KN_3$ can induce cg-N. Furthermore, by grinding or loading pressure on the mixture of $KN_3$ with Li, Na, K, Cs, Ca, Mg and Al, we find that elements with higher electronegativity, higher pressure and temperature conditions are needed to induce cg-N, while grinding alone is sufficient for alkali metals even without heating, thus confirming the charge-transfer mechanism. These findings provide guidance for the synthesis of cg-N under ambient conditions through metal-catalyzed polymerization of various azides.

## INTRODUCTION

Cubic gauche nitrogen (cg-N) with similar structure to diamond is a representative pure nitrogen polymer, characterized by a three-dimensional network of single N-N bonds.[1-15] Due to the significant energy difference between N–N single bonds (160 kJ/mol) and N≡N triple bonds (954 kJ/mol), cg-N possesses high-energy density and offers the potential for clean energy release upon depolymerization[1-3,16-19]. In 1992, the cg-N was theoretically predicted to be stable at high pressures, identifying it as a cubic structure with the space group $I2_13$[2]. The experimental validation of cg-N's structure was achieved by using diamond anvil cell (DAC) under conditions exceeding 110 GPa, and distinct N–N Raman modes was observed for the first time[16]. However, the synthesis of cg-N at such extreme pressures and its limited stability at ambient conditions present significant barriers to its practical applications[16,20-22].

While theoretical studies have suggested that cg-N can be stable at ambient pressure, experimental realization under these conditions has been challenging[3,20,22-25]. Researchers tried several routes to synthesize cg-N at atmospheric pressure by using azides as precursor[26-31]. Trace amounts of cg-N were synthesized at atmospheric pressure using highly toxic and sensitive sodium azide ($NaN_3$) precursor based on plasma-enhanced chemical vapor deposition (PECVD) method, but the conversion rate needs to be improved through carbon nanotube confinement effect[26]. Based on the PECVD, free-standing cg-N stable at 760 K was obtained by using potassium azide ($KN_3$) as precursor, in which a cg-N content of ~0.7 wt% was realized and rapid and intense thermal decomposition behavior was also observed[27]. Very recently, cg-N has been successfully synthesized by simple one-step heating of azide under vacuum or protective atmospheres without the need for PECVD, giving a higher cg-N content of ~1.5 wt%[31]. These discoveries are expected to generate significant interest in the practical applications of cg-N.

At present, the main challenge hindering the practical application of cg-N is its low synthesis efficiency. Although, the highest reported cg-N content of ~1.5 wt%, achieved through the heating of potassium azide[31], is twice the content (~0.7 wt %) obtained using PECVD techniques[27,31], this yield remains insufficient for the practical applications of cg-N. The primary reason for this is the lack of a comprehensive understanding of the synthesis mechanism. Therefore, elucidating the polymerization mechanism of nitrogen at ambient pressure is crucial for developing more efficient production methods for cg-N.

Earlier studies have suggested that surface instability is a key factor contributing to the decomposition of polymerized cg-N[23]. Strategies such as hydrogen saturation have shown promise in stabilizing cg-N surfaces by saturating dangling bonds and facilitating charge transfer[23]. These theoretical findings provide new insights, indicating that charge transfer not only stabilizes the surface of cg-N but also plays a critical role in enabling its synthesis under ambient conditions. In fact, charge transfer through chemical doping has traditionally been a preferred method for achieving nitrogen polymerization. For instance, the introduction of metals has been shown to significantly reduce synthesis pressures[32-43], as exemplified by the synthesis of $LiN_5$ at 45 GPa[32] and $K_2N_6$ at 50 GPa[35].

Based on these insights, we first investigated the synthesis of $KN_3$ under various atmospheres to identify the key factors controlling the formation of cg-N at ambient pressure. Our findings suggest that in a dry environment, the potassium (K) element released during the thermal decomposition of $KN_3$ plays a crucial role in the polymerization of N, which is suppressed in a moisture case. We then explored the use of various metals with different electronegativities for the synthesis of cg-N in place of potassium. We discovered that metals with high electronegativity can induce the polymerization of N at ambient pressure and room temperature through surface absorption. These findings suggest the charge transfer from metal to nitrogen triggers the polymerization of nitrogen at ambient conditions, clarifying the long puzzle of the mechanism for synthesizing cg-N from azide.

**RESULTS**

**A Moisture hinders the synthesis of cg-N through heating azides**

Previous studies have demonstrated that cg-N can be synthesized from $KN_3$ through heating under vacuum or protective atmospheres. However, when $KN_3$ is heated in air, cg-N formation is not observed[31]. To elucidate the polymerization mechanism, it is crucial to first identify the key role of air in preventing the formation of cg-N. We attempted to synthesize cg-N by heating $KN_3$ in various gases, including pure $N_2$, pure $O_2$, and pure $CO_2$, which are the main components of air. As illustrated in Fig. 1, subsequent to heat treatment under all these conditions, the sample transitions from white to green color. The Raman spectra indicate the formation of cg-N, characterized by the modes at 640 $cm^{-1}$ in all green samples, suggesting these the main components of air do not influence the polymerization of nitrogen.

We then investigated the role of moisture in the synthesis of cg-N. As shown in Fig. 1, upon heating $KN_3$ in dry air, the sample color transitions to green, and characteristic Raman modes indicative of cg-N formation are detected. To further substantiate the inhibitory effect of moisture on cg-N synthesis, experiments were conducted in which $KN_3$ and water were co-heated in an $N_2$ protective atmosphere. Importantly, $KN_3$ was not in direct contact with water, ensuring that any observed effects were attributable solely to the presence of water vapor rather than direct interaction between the two. In this scenario, the sample remains unreacted, as illustrated in Fig. 1. These findings strongly illustrate that the moisture can prevent the transition from $KN_3$ to cg-N.

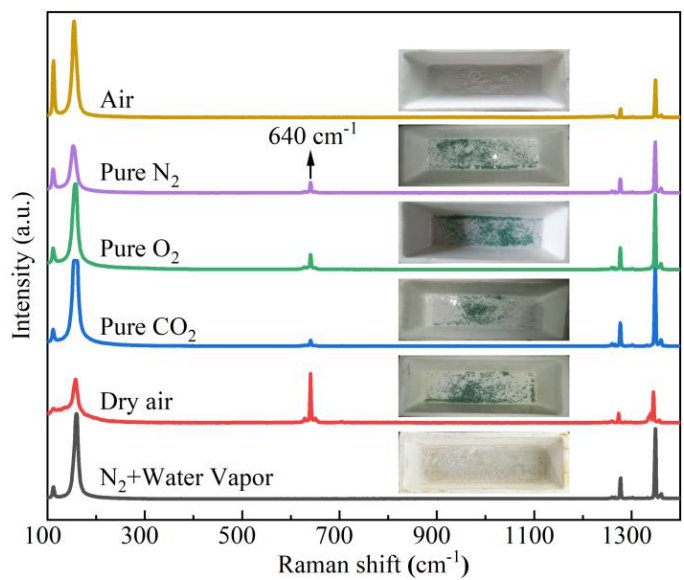

**Fig. 1 | Raman spectra of KN$_3$ powder after heat treatment under various atmospheric conditions.** The experiments involved heating pure KN$_3$ powder in a controlled treatment chamber under different atmospheres, including air, dry air, pure N$_2$, pure O$_2$, pure CO$_2$, and N$_2$ with water vapor. The inset presents a digital image of the samples following heat treatment under each respective condition.

Interestingly, a detailed examination of the product after heating KN$_3$ in air reveals that a few micrometer-sized particles exhibit a distinct color compared to the unreacted KN$_3$, as shown in Fig. 2. An enlarged view of these particles in Fig. 2b displays a green color, similar to that of cg-N in Fig. 2a. Employing focused Raman spectroscopy, as shown in Fig. 2c, a mode at 640 cm$^{-1}$ confirms that these green particles are indeed cg-N, contrary to earlier reports suggesting that cg-N cannot be synthesized in air. We compared the cg-N particles synthesized in air and in N$_2$. The cg-N particles synthesized in N$_2$ exhibit clean surfaces, whereas those synthesized in air are encapsulated within transparent KN$_3$. These observations suggest that encapsulation isolates the KN$_3$ from air, preventing reactions with water during heating. Consequently, only a minimal amount of cg-N forms from KN$_3$ when heated in air.

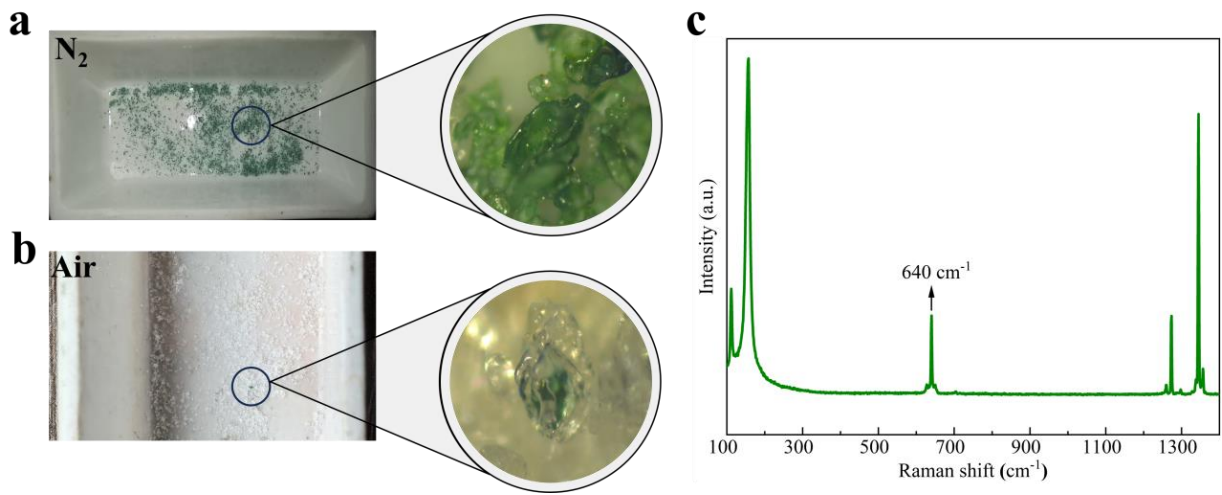

**Fig. 2 | cg-N synthesized in air. a**, cg-N synthesized in N$_2$ protective gas, with magnified detail. **b**, cg-N synthesized in air protective gas, with magnified detail. **c**, Raman spectra of the cg-N synthesized in air protective gas.

**B Gas chromatographic analysis confirms the existence of decomposed K**

If decomposed K play a key role on the synthesizing of cg-N, some unreacted metal K should remain in the product after heating. To verify the composition of the product, we conducted gas chromatographic analysis on the sample dissolved in water. A small amount of KN$_3$ sample heated in N$_2$, O$_2$, CO$_2$ and air protective gas are added in deionized water respectively. KN$_3$ sample heated in air (Supplementary Fig. S1a) dissolves immediately and no bubble appear. In the case of KN$_3$ sample heated in N$_2$, O$_2$, and CO$_2$ protective gas, a large number of bubbles were generated. For demonstration purposes, the KN$_3$ sample heated in N$_2$ was selected to illustrate the bubble formation (Supplementary Fig. S1a). To analyze the composition of the bubbles, using gas chromatographic analysis the H$_2$, N$_2$ and O$_2$ can be detected in the generated gas (Supplementary Fig. S1b). Considering the possible elements in the powder sample, H$_2$ can only be generated from the reaction between metallic potassium and water. Thus, gas chromatographic analysis confirms the existence of decomposed K when cg-N is synthesized from KN$_3$. These findings indicate that the decomposition of K from KN₃ is pivotal in the polymerization of nitrogen, and water inhibits the synthesis of cg-N by preventing the decomposition of KN₃ during heating.

**C Synthesizing cg-N without heating through Charge transfer**

Our previous research indicates that potassium can stabilize the interface of cg-N, thereby enhancing its stability via charge transfer[23,27,31]. Earlier studies have also demonstrated that metal doping can reduce the polymerization pressure of nitrogen through similar charge transfer mechanisms[32-43]. Based on these findings, we hypothesize that potassium acts as a catalyst in the transition from KN₃ to cg-N and enhances its stability under ambient conditions through charge transfer processes.

Based on the above analysis of the key role of K in synthesizing cg-N, we designed the following experiment to further verify our assumption of charge transfer mechanism on the polymerization of N. If the decomposed K transfers charge to the azide ion and induces the formation of cg-N, we consider adding additional metal K to azides could promote the polymerization process. Thus, we mixed powder potassium azide and metal potassium, and found the formation of cg-N compound in the grinding mixture even without heating, which confirms that charge transfer plays a key role in the formation of polymerized nitrogen from azides. The experimental details are as follows.

**Raman spectroscopy analysis.** By placing powdered potassium azide and metallic potassium in a mortar and grinding in the argon atmosphere of the glove box, we found that the formation of cg-N compound in the grinding mixture using shear stress. In Fig. 3a, the characteristic peak at 1269 cm⁻¹ corresponds to the $2\nu_2$ mode of the azide ion[44]. Compare with the Raman mode of pure KN₃, new lines observed in the mixed KN₃ and K compound, including an intense line at 638 cm⁻¹ with a weaker shoulder near 625 cm⁻¹, a new line at 703 cm⁻¹ with a factor of ~25 lower intensity than the line at 638 cm$^{-1}$, which was not only in good agreement with theoretical calculations about the Raman modes of cg-N under zero pressure[21], but also the experiment results of cg-N synthesized by plasma treatment methods[26,27]. So in the Raman spectra of our experiment (Fig. 3a), the prominent line at 638 cm$^{-1}$ and the weaker line at 703 cm$^{-1}$ can be definitively attributed to the pore-breathing A symmetry and N-N tilting T(TO) symmetry Raman-active modes, respectively, of cg-N extrapolated to near ambient pressure, which strongly indicates the formation of cg-N in the compound.

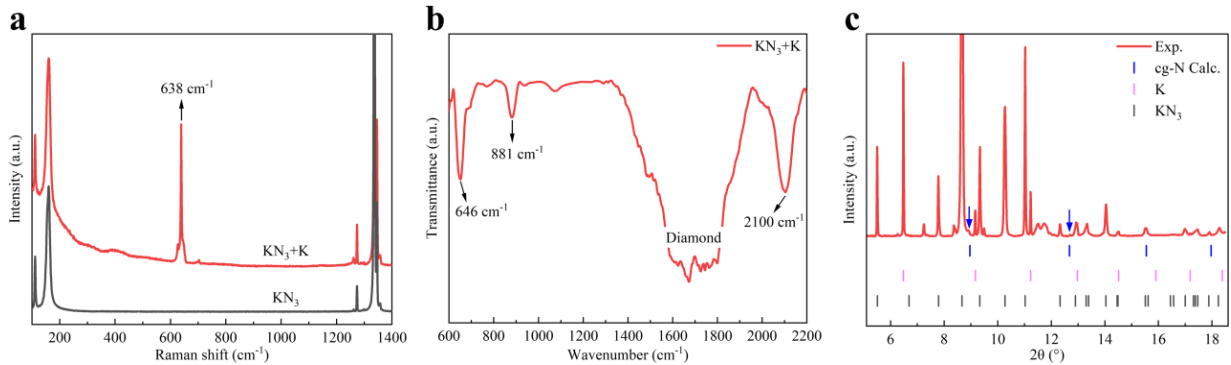

**Fig. 3 | Spectroscopic characterization of the cg-N compound. a**, Raman spectra of pure KN$_3$ powder and compound contains KN$_3$ and K. **b**, FTIR spectra of the cg-N compound at 0.1 GPa. **c**, Diffraction patterns measured at 0.2 GPa for the compound using synchrotron radiation X-rays with a wavelength of 0.4142 Å; peaks marked with arrows correspond to simulated X-ray diffraction peaks of cg-N at 0.2 GPa.

**FTIR spectroscopy analysis**. The collection of infrared spectral signals at atmospheric pressure is difficult due to the influence of air. Considering the reactivity of metallic potassium, we chose to use DAC to package the mixture of KN$_3$ and K at a pressure of 0.1 GPa, and obtained the infrared characteristic signal of cg-N. New absorptions at 881 cm$^{-1}$ correlate with the T(TO) symmetry vibration of cg-N extrapolated to near zero pressure in agreement with the calculation of Caracas[21], while absorption peaks at 646 and 2100 cm$^{-1}$ arise from bending $\nu_2$ mode and the antisymmetric stretching $\nu_3$ mode of the azide ion of unreacted $KN_3$[26,27], confirming partial azide preservation (Fig. 3b). Moreover, the broad IR modes at approximately 1500 to 1800 cm$^{-1}$ are due to the diamond.

**X-ray powder diffraction.** To elucidate the chemical structure of this compound, we conducted X-ray powder diffraction (XRD) measurements of the compound at 0.2 GPa. Fig. 3c displays the XRD pattern obtained at 0.2 GPa using synchrotron radiation with a wavelength of 0.4142 Å. Despite strong diffraction peaks of $KN_3$ overshadowing cg-N signals, the XRD spectra successfully captured distinct diffraction signals for cg-N at 0.2 GPa. By comparing the observed patterns with simulated cg-N diffraction data, we identified two weak but discernible peaks in Fig. 3c (marked by arrows) that align with the cg-N structure. We also obtained the XRD pattern of cg-N compound at ambient pressure (Supplementary Fig. S2) using with CuKα radiation ($\lambda \approx 1.54$ Å), it was identified a weak but discernible peak near 80° in that align with simulated cg-N diffraction data at ambient pressure, providing additional confirmation of the presence of cg-N within the compound.

### D. Electronegativity dominated synthesizing conditions of cg-N

Considering that the charge transfer between elements is determined by electronegativity, various metals with increased electronegativity were selected to investigate their influence on the polymerization of nitrogen. The results of lithium (Li), sodium (Na), potassium (K), cesium (Cs), calcium (Ca), magnesium (Mg) and aluminum (Al) metals with the electronegativity from 0.8 to 1.6 are shown in Fig. 4.

In the synthesis experiments for cg-N, similar Raman features were observed when using other alkali metals, such as Li, Na, and Cs, combined with $KN_3$ as precursors. This strongly supports the significant role that metals play in facilitating cg-N formation. For metals with higher electronegativity, grinding alone is insufficient to effectively induce reactivity. However, combining high-pressure and high-temperature conditions can yield similar outcomes. The mixtures of Ca with KN3 under these conditions produce the characteristic Raman signals of cg-N. For Mg and Al, which achieve Raman signal of N-N single bond at higher pressure and temperature. Importantly, cg-N synthesized with the participation of different metals can be stable under environmental pressure. Notably, metals with higher electronegativity exhibit greater inertia and require more energy input to activate the formation of cg-N.

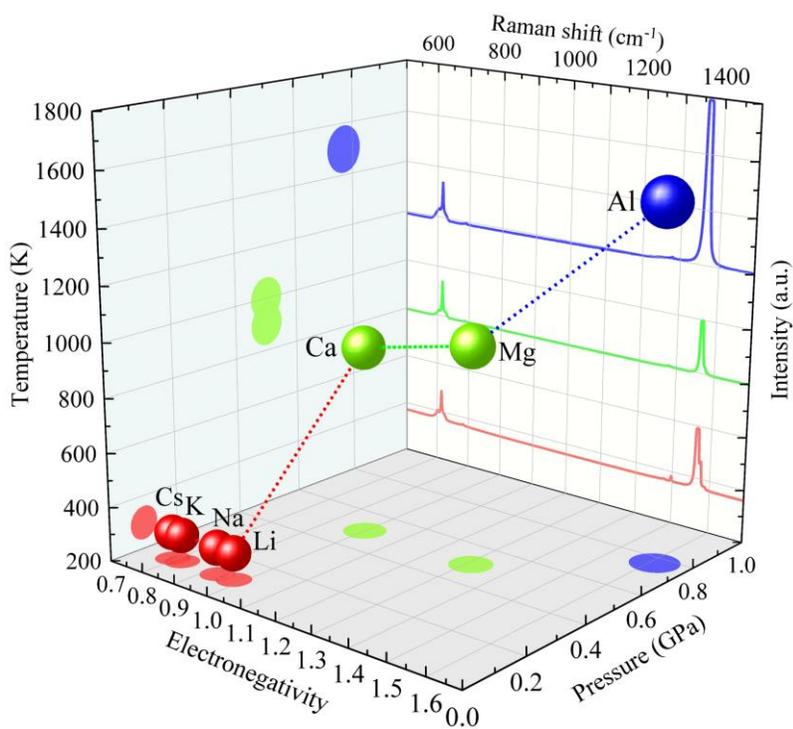

**Fig. 4 | Synthesis conditions for cg-N compounds from metals with varying electronegativity.** Selected Raman spectra of cg-N compounds are shown on a white panel, with red, green, and blue curves representing Li, Ca, and Al, respectively.

## DISCUSSION

By Raman spectroscopy and gas chromatography analysis of potassium azide after heat treatment in different atmospheres, we confirmed the presence of decomposed potassium (K) during synthesis and demonstrated that water inhibits the formation of cubic gauche nitrogen (cg-N) in air by preventing the decomposition of $KN_3$. This finding underscores the importance of controlling the synthetic environment to exclude moisture, which would disrupt the role of K as a charge donor (details analysis can be found in the SI files). Additionally, grinding experiments with potassium metal and potassium azide revealed that the addition of potassium metal to $KN_3$ can also induce the formation of cg-N. Raman spectroscopy, FTIR spectra, and X-ray diffraction analysis provide strong evidence for the formation of cg-N, which is consistent with both theoretical predictions and experimental results from plasma-assisted synthesis methods. Expanding on this, we investigated the effects of metals with varying electronegativity on cg-N synthesis. Less electronegative metals, such as alkali metals, effectively promote cg-N formation through grinding. In contrast, metals with higher electronegativity require elevated temperatures and pressures to activate the nitrogen polymerization process. This electronegativity-dependent behavior highlights the critical role of charge transfer in promoting cg-N and suggests that cg-N synthesis could be achieved under ambient conditions using mechanical methods such as grinding. Similar phenomena were also observed in sodium azide, while the signal is relatively weak. These findings highlight the potential to optimize synthesis conditions by leveraging charge transfer from azides, offering a transformative approach to cg-N synthesis under environmental conditions.

Upon elucidating the polymerization mechanism of cg-N, we have derived guidelines for the synthesis of cg-N from azides under ambient conditions. Firstly, the use of catalysts with low electronegativity, such as alkali metals, is essential to initiate polymerization via charge transfer. These catalysts can be introduced either by mixing the azides with metal particles or through the decomposition of the azides themselves. Secondly, meticulous care must be taken to protect the reaction from moisture, as catalysts with low electronegativity are highly sensitive to water. This can be achieved through the use of protective gases, vacuum environments, encapsulation techniques, etc.

Furthermore, our findings provide insights into strategies for enhancing the yield of cg-N in future syntheses. Currently, the synthesis of cg-N from azides under ambient conditions occurs exclusively at the surface of the azide particles, leading to low yields. Even in heating methods, decomposition may initially occur at the surface of the azide particles, followed by the reaction of the decomposed metal with azides. To enhance efficiency, one can increase the specific surface area of azides using various nanotechnologies, such as reducing particle size, employing porous structures, or fabricating films or nano-sheets.

## CONCLUSSION

This study identifies charge transfer as the key mechanism driving the synthesis of cubic gauche cg-N from azides under ambient conditions. We demonstrate that potassium, released during the thermal decomposition of $KN_3$ in controlled, moisture-free atmospheres, plays a pivotal role in inducing nitrogen polymerization. Our results show that the presence of water molecules inhibits the decomposition of $KN_3$, thereby preventing charge transfer induced cg-N. Additionally, we find that alkali metals such as lithium, sodium, and potassium facilitate cg-N synthesis at ambient pressure and temperature, while metals with higher electronegativity require elevated conditions to trigger polymerization. These findings highlight the critical role of charge transfer in cg-N formation. This work advances our understanding of the cg-N synthesis mechanism and provides practical guidelines for optimizing synthesis conditions. The discovery that charge transfer enables nitrogen polymerization under mild conditions offers a promising pathway for scalable production of cg-N and other high-energy nitrogen-rich materials.